\documentclass[aps,prc,twocolumn,floatfix,nofootinbib,preprintnumbers,superscriptaddress,longbibliography]{revtex4-1}

\usepackage{CJKutf8}
\usepackage{epsfig}
\usepackage{color}
\usepackage{rotating}
\usepackage{natbib}
\usepackage{verbatim}
\usepackage{epstopdf}
\usepackage{graphicx}
\usepackage{changes}
\usepackage{bm}
\usepackage{hyperref}
\usepackage{multirow}
\usepackage{makecell}
\usepackage{enumitem}
\usepackage{amsmath}

\usepackage{array}
\usepackage{longtable}
\usepackage{tabularx}
\usepackage[para,online,flushleft]{threeparttablex}
\usepackage{array} 
\usepackage{listings}
\usepackage[para,online,flushleft]{threeparttablex}
\usepackage{tabularx}
\usepackage{booktabs,dcolumn}
\usepackage{textpos}
\usepackage{booktabs}
\usepackage{multirow,bigdelim}
\usepackage{float}
\usepackage{diagbox}
\newcolumntype{C}[1]{>{\centering\let\newline\\\arraybackslash\hspace{0pt}}m{#1}}

\graphicspath{{./figures/}}

\makeatletter
\def\@bibdataout@aps{%
\immediate\write\@bibdataout{%
@CONTROL{%
apsrev41Control%
\longbibliography@sw{%
    ,author="08",editor="1",pages="1",title="0",year="1"%
    }{%
    ,author="08",editor="1",pages="1",title="",year="1"%
    }%
  }%
}%
\if@filesw \immediate \write \@auxout {\string \citation {apsrev41Control}}\fi 
}
\makeatother

\newcommand{\soft}[1]{\texttt{#1 }}
\newcommand{\lisepp}{\soft{LISE$^{++}$}}

\begin{document}

\begin{CJK*}{UTF8}{gbsn}

\title{Quantified limits of the nuclear landscape}

\author{L{\'e}o Neufcourt}
\affiliation{Department of Statistics and Probability, Michigan State University, East Lansing, Michigan 48824, USA}
\affiliation{Facility for Rare Isotope Beams, Michigan State University, East Lansing, Michigan 48824, USA}

\author{Yuchen Cao (曹宇晨)}
\affiliation{Facility for Rare Isotope Beams, Michigan State University, East Lansing, Michigan 48824, USA}
\affiliation{National Superconducting Cyclotron Laboratory, Michigan State University, East Lansing, Michigan 48824, USA}

\author{Samuel A. Giuliani}
\affiliation{Facility for Rare Isotope Beams, Michigan State University, East Lansing, Michigan 48824, USA}
\affiliation{National Superconducting Cyclotron Laboratory, Michigan State University, East Lansing, Michigan 48824, USA}

\author{Witold Nazarewicz}
\affiliation{Facility for Rare Isotope Beams, Michigan State University, East Lansing, Michigan 48824, USA}
\affiliation{Department of Physics and Astronomy, Michigan State University, East Lansing, Michigan 48824, USA}

\author{Erik Olsen}
\affiliation{Institut d'Astronomie et d'Astrophysique, 
Universit{\'e} Libre de Bruxelles, 1050 Brussels, Belgium}

\author{Oleg B. Tarasov}
\affiliation{National Superconducting Cyclotron Laboratory, Michigan State University, East Lansing, Michigan 48824, USA}

\date{\today}

\begin{abstract}
\begin{description}

\item[Background] 
The chart of the nuclides is limited by particle drip lines beyond which
 nuclear stability to proton or neutron  emission  is lost.
Predicting the range  of particle-bound isotopes 
poses an appreciable challenge for nuclear theory as it involves 
extreme extrapolations of nuclear masses well beyond the  regions where experimental information is available. Still, quantified extrapolations are 
crucial for a wide variety of  applications, 
including the modeling of stellar nucleosynthesis.

\item[Purpose]
We use microscopic nuclear global mass models, current mass data,  and Bayesian methodology to provide quantified predictions of proton and neutron separation energies  as well as Bayesian probabilities of existence throughout the nuclear landscape all the way to the particle drip lines. 

\item[Methods]
We apply nuclear density functional theory with several energy density functionals. We also consider two global mass models often used in astrophysical nucleosynthesis 
simulations.  To account for uncertainties, Bayesian Gaussian processes are trained on the separation-energy residuals for each individual model, and the resulting  predictions are  combined via Bayesian model averaging. 
This framework allows to account for
systematic and statistical uncertainties and propagate them to 
extrapolative predictions.

\item[Results]
We establish and characterize the drip-line regions where the probability that the nucleus is particle-bound decreases from $1$ to $0$. In these regions, we provide quantified predictions for one- and two-nucleon separation energies.
According to our Bayesian model averaging analysis,
7759 nuclei with $Z\leq 119$ have a  probability of existence $\geq 0.5$. 

\item[Conclusions]
The extrapolation results obtained in this study
will be  put through stringent tests
when new experimental information on existence and masses of exotic nuclei
 becomes available. In this respect, the  quantified landscape of nuclear existence obtained in this study should be viewed as a dynamical prediction that will be fine-tuned when new experimental information  and improved global mass models become available.

\end{description}
\end{abstract}

\maketitle
\end{CJK*}

\section{Introduction} 

Of the several thousand atomic nuclei thought to exist, only around 3,000 have been experimentally observed, and only 286 are considered to be primordial nuclides (i.e., isotopes  found on  Earth that have existed in their current form since before Earth was formed). All nuclear species can be mapped on the chart of nuclides, or nuclear landscape. The landscape's boundaries, the particle drip lines,  mark the end of nuclear binding. 
On the proton-rich side, the drip line has been reached experimentally all the way up to $_{93}$Np \cite{Zhang2019}. On the other hand, the neutron drip line  has been delineated only for light nuclei up to $_{10}$Ne \cite{Ahn2019} and
for  heavier elements it is based on  theoretical predictions.

Quantifying the limits of nuclear binding is important for understanding the origin of elements in the universe. In particular, the astrophysical rapid neutron capture \mbox{(r-)} process responsible for the generation of many heavy elements is believed to operate very closely to the  neutron drip line; hence, the structure of very exotic nuclei directly impacts the way the elements are produced in stellar nucleosynthesis
\cite{Horowitz2019}.
A quantitative understanding of the r-process requires knowledge of 
nuclear properties and reaction rates of $\sim$3,000  very neutron-rich isotopes, many of which cannot be reached experimentally. The missing nuclear data for astrophysical simulations must be provided by massive extrapolations based on nuclear models augmented by the most recent experimental data. Here, Bayesian machine learning, with its unified statistical treatment of all uncertainties, is the tool of choice when aiming at  informed predictions including both a reduction of extrapolation errors and quantified bounds.

The global modeling of all nuclei, including complex exotic nuclei far from stability, is a challenging quest that  requires  control of many aspects of the nuclear many-body problem. For such a task, the microscopic tool of choice is  nuclear density functional theory (DFT) rooted in the mean-field approach \cite{Ben03}.  During recent years, several global DFT mass tables have been calculated using different energy density functionals (EDFs): Skyrme \cite{Erl12a,Erler13,Wang2015}, Gogny \cite{Goriely09,Delaroche2010}, and covariant \cite{Afanasjev13,Agbemava2014,Wang2015,Xia17}.
Other well-calibrated mass models include
the microscopic–macroscopic finite-range droplet model (FRDM) \cite{Moller2012} and  Skyrme–HFB models based on the Hartree–Fock–Bogoliubov (HFB) method \cite{Goriely2013}. 
The number of predicted bound nuclides with atomic numbers between 2 and 120 shows significant model variations:
for instance,  it is around 7,000 in the Skyrme-DFT analysis	\cite{Erl12a} while it is over 9,000 in the covariant DFT approach of Ref.~\cite{Xia17}.

The systematic uncertainty on masses has often been estimated by an analysis of intermodel dependencies through  comparing predictions of different DFT frameworks and different EDF parametrizations \cite{Erl12a,Afanasjev15,Wang2015}.
Statistical uncertainties are best evaluated by means of Bayesian inference methods involving full parameter estimation \cite{McDonnell2015}. The uncertainties on calculated masses impact nuclear astrophysical calculations
such as the r-process abundance predictions \cite{Martin2016,Mumpower2016,Sprouse2019,Horowitz2019}. To  improve
the quality of theoretical mass predictions and 
minimize  uncertainties, diverse machine learning techniques have been applied
\cite{Athanassopoulos2004,Utama16,Utama17,Utama18,Zhang2017,Niu2018,Neufcourt2018} that combine theoretical modeling with 
currently available experimental information.

In this study, we combine results of several global mass models and information contained in experimental masses to make
a quantified assessment of proton and neutron separation energies and drip lines. To this end, we employ the technique of
Bayesian model averaging (BMA)
\cite{Hoeting1999,Was00,Bernardo1994}, which has recently been  adopted to provide quantified predictions for both neutron-rich nuclei in the Ca region
\cite{Neufcourt2019a} and two-proton emitters \cite{Neufcourt2020}.

The paper is organized as follows. Section~\ref{methods} describes  the nuclear mass models used  and the statistical methodology adopted in our work.
The results obtained in this study are discussed in Sec. \ref{results}. Finally, Sec.~\ref{conclusions} contains a summary and conclusions.
The tables of  separation energies (with uncertainties) predicted in our BMA calculations are provided in the Supplemental Material \cite{SM} together with downloadable  plots of the quantified landscape of nuclear existence (Fig.~\ref{fig:posterior-landscape}) 
and the quantified  separation energy landscape in the neutron drip line region (Fig.~\ref{fig:neutron-dripline}) in PDF format.

\section{Methods}\label{methods}

\subsection{Nuclear mass models}

In our study, we consider 8 models based on nuclear DFT: the Skyrme energy density functionals
SkM$^*$ \cite{Bartel1982}, SkP \cite{Dob84}, SLy4 \cite{Chabanat1995}, SV-min \cite{Kluepfel2009}, UNEDF0 \cite{UNEDF0},  UNEDF1 \cite{UNEDF1}, and  UNEDF2 \cite{UNEDF2} as well as the Gogny functional D1M~\cite{Goriely09} and 
the functional BCPM \cite{Baldo13}. For each model, the mass table of even-even nuclei was computed self-consistently by solving the Hartree-Fock-Bogoliubov (HFB) equations as described in Refs.~\cite{Erl12a,Neufcourt2018,Neufcourt2019a,Neufcourt2020};
masses of odd-$Z$ and odd-$N$ systems were then extracted using computed pairing gaps \cite{Neufcourt2019a,massexplorer}.

The above set  of DFT models was augmented by two mass-optimized  mass  models commonly used in nuclear astrophysics studies: the microscopic-macroscopic  model FRDM-2012 \cite{Moller2012} and the Skyrme-HFB model HFB-24 \cite{Goriely2013}. 

The above models were optimized using different strategies and varied datasets involving global nuclear observables and, sometimes, pseudodata such as  nuclear matter parameters \cite{Ben03}. Consequently, the accuracy of these models with respect to measured masses (measured in terms of  the root-mean-square (rms)  deviation) varies between several MeV (SkM$^*$) and $\sim$600\,keV (FRDM-2012 and HFB-24) \cite{Horowitz2019}. 
Still, the rms mass deviations are reduced to similar values
across models following statistical treatment,
 as demonstrated in 
Refs.~\cite{Neufcourt2018,Neufcourt2020}.

\subsection{Statistical methods}

Our methodology follows closely our previous work \cite{Neufcourt2018,Neufcourt2019a,Neufcourt2020} in which we combined the current theoretical and experimental information using Bayesian simulations to arrive at  informed predictions.

\subsubsection{Gaussian processes} 

The Bayesian statistical model for separation-energy residuals,
i.e., differences $y_i= y^{exp}(x_i) - y^{th}(x_i)$  between experimental data and theoretical predictions,
 can be written as:
\begin{equation}
y_{i}= f(x_i, \theta)+\sigma\epsilon_{i},
\end{equation}
where the function $f(x,\theta)$ represents the systematic deviation 
and $\sigma \epsilon$  is the propagated statistical uncertainty.

Quantified extrapolations $y^*$   are obtained from the posterior predictive
 distribution $p(y^*|y)$ using a stationary Markov chain.
Similarly to our previous studies, we model independently 
$S_{1n}$, $S_{2n}$, $S_{1p}$ and $S_{2p}$ on the  four subsets of nuclei defined by the particle-number parities (even-even, even-odd, etc.).
By doing this we are ignoring some (slight) correlations 
between systematic uncertainties.

For the function $f$ we take a Gaussian process on the two-dimensional nuclear domain indexed by $x=(Z, N)$, characterized by its mean $\mu$ 
(taken here as a scalar parameter) and covariance $k$:
\begin{equation}\label{GPmodel}
f(x,\theta)\sim\mathcal{GP}(\mu, k_{\eta,\rho}(x, x')).
\end{equation}
The ``spatial" dependence between nearby nuclei 
is represented by an exponential quadratic covariance kernel: 
\begin{equation}\label{kernel}
k_{\eta ,\rho }(x,x^{\prime }):=\eta ^{2}e^{-\frac{(Z-Z^{\prime })^{2}}{%
2\rho _{Z}^{2}}-\frac{(N-N^{\prime })^{2}}{2\rho _{N}^{2}}},
\end{equation}
where the parameters $\eta$, $\rho_{Z}$ and $\rho_{N}$ 
represent, respectively, the scale and characteristic  correlation ranges  in the proton and neutron directions.
Consequently,  our statistical model has four parameters
 $\theta:=(\mu,\eta,\rho_Z,\rho_N)$. 
We have found in a previous study \cite{Neufcourt2018} 
that Gaussian processes overall outperform Bayesian neural networks, achieving similar rms deviations 
with a more faithful uncertainty quantification and considerably fewer parameters. We have also demonstrated \cite{Neufcourt2019a,Neufcourt2020} that the parameters $\theta$  are  well constrained and fairly uncorrelated. It is worth noting that
a  non-zero value of the GP mean prediction $\mu$ allows to reproduce more consistently the extrapolative data. This  GP extension to  nonzero $\mu$ \cite{Neufcourt2020} significantly improves results.

\begin{table*}[htb!]
	\caption{Model posterior weights obtained in {the} variants  BMA($n$) (\ref{eq:neutron-weights}) and  BMA($p$) (\ref{eq:proton-weights}) of our BMA calculations.   For compactness, the following abbreviations are used: UNEn=UNEDFn (n=0,1,2) and FRDM=FRDM-2012.}
  \label{table:post-weights}%
      \begin{ruledtabular}
    \begin{tabular}{c|ccccccccccc}
BMA variant & {SkM*} & {SkP} & {SLy4} & {SV-min} & {UNE0} & {UNE1} & {UNE2} & {BCPM} & {D1M} & {FRDM} & {HFB-24} \\ \hline
BMA($n$) & 0.10 & 0.10 & 0.06 & 0.11 & 0.12 & 0.10 & 0.09 & 0.06 & 0.04 & 0.12  & 0.09 \\
BMA($p$) & 0.00 & 0.03 & 0.08 & 0.05 & 0.04 & 0.14 & 0.12 & 0.04 & 0.16  & 0.17 & 0.17
    \end{tabular}%
    \end{ruledtabular}
\end{table*}

\subsubsection{Datasets}

Our dataset combines all experimental masses from
AME2003 \cite{AME03b} and AME2016 \cite{AME16b} 
augmented by the recently measured  masses from Refs.~\cite{deRoubin17,Welker17,JYFLTRAP,Leistenschneider2018,Michimasa2018,Oxford18,Ito18}.
For nuclei where experiments have been repeated, 
we take the most recent value. For testing purposes we split this dataset into a training set (AME2003) 
and a testing set 
(AME16-03: all masses in AME2016+ that are not in AME2003),

For prediction purposes,  we use 
the full mass dataset for training
-- the performance of the statistical model was assessed in previous work
\cite{Neufcourt2018,Neufcourt2019a,Neufcourt2020} --
and carry out  calculations based on a  large set of nuclei 
for which raw theoretical separation energies are not too negative;
this includes all  proton-bound nuclei.
Nuclei with negative experimental separation energies, e.g., narrow ground-state proton resonances,  {have not been} used for training.

\begin{figure*}[htb!]
\includegraphics[width=0.9\linewidth]{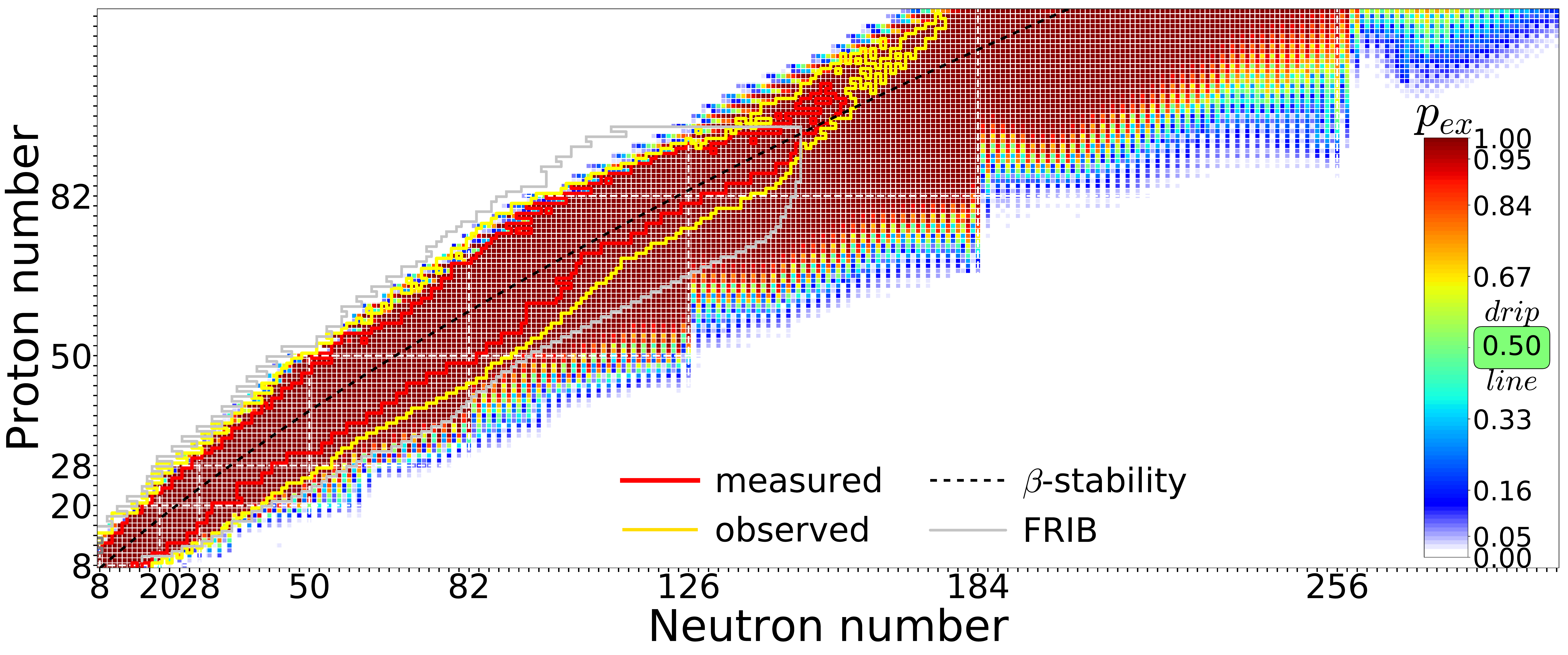}
\caption{
The quantified  landscape of nuclear existence obtained in our BMA calculations. For every nucleus with $Z,N \ge 8$ and $Z\le 119$
 the probability of existence $p_{\rm ex}$ (\ref{pex}),
 i.e., the probability that  the nucleus is bound with respect to proton and neutron decay, is marked.
The domains of nuclei which have been experimentally observed and whose separation energies have been  measured (and  used for training) are indicated.
To provide a realistic estimate of the discovery potential with  modern radioactive ion-beam facilities,
the isotopes within  FRIB's experimental  reach are marked. The magic numbers are shown by straight (white) dashed lines, and  the average line of $\beta$-stability  defined as in Ref.~\cite{Kodama1971}
is marked by a  (black) dashed line.  See text for details.
This figure (without the FRIB range), in PDF format, can be downloaded from \cite{SM}.
}
\label{fig:posterior-landscape}
\end{figure*}

\subsubsection{Computations}

Samples from posterior distributions were obtained 
from 50,000 iterations of Monte Carlo Markov chains, 
after the stationary state was reached 
(with 50,000 samples previously burnt-in), 
which were used in turn to generate 10,000 mass tables.

\subsubsection{Bayesian model averaging}

 Based on general considerations \cite{Kej19}, it is expected that BMA should on average outperform individual models.
Similarly as in Ref.~\cite{Neufcourt2019a,Neufcourt2020}, also
dealing with  model-based extrapolations, in this study
we employ the BMA framework  to select the models with the best predictive power and  avoid overfitting. 
This brings us to the  BMA variants developed in our previous studies
where the weight of each model $\mathcal{M}_k$ 
is based on its capacity  to account for known experimental data 
at the exterior of the training dataset.

We use two families of weights based on the data from the neutron-rich and proton-rich nuclear domains.
On the neutron-rich side, we follow Ref.~\cite{Neufcourt2019a} and use the weights
\begin{equation}\label{eq:neutron-weights}
w_k(n) :\propto p\left(S_{1n/2n}(x)>0|\mathcal{M}_k \text{ for } x\in\mathcal{D}_n\right),
\end{equation}
where $\mathcal{D}_n$ is the set of 254 experimentally observed neutron-rich nuclei  with $20\le Z \le  50$
for  which no  experimental neutron separation energy is available (such as $^{60}$Ca \cite{Tarasov2018}).
On the proton-rich side, the experimental reach goes beyond the proton drip line, as separation energies have been established experimentally  for many one- and two-proton emitters.  
To this end, in this region we follow Ref.~\cite{Neufcourt2020} and
use the weights given by
\begin{equation}\label{eq:proton-weights}
w_k(p) :\propto p\left(Q_{2p}(x)>0, S_{1p}(x) > 0|\mathcal{M}_k \text{ for } x\in\mathcal{X}_{2p}\right), 
\end{equation}
where $\mathcal{X}_{2p}$ is the set of five long-lived two-proton emitters 
$^{19}$Mg, 
$^{45}$Fe, $^{48}$Ni,  $^{54}$Zn,
and  $^{67}$Kr (see Ref.~\cite{Neufcourt2020} for more discussion). In the following, we
 refer to these variants as BMA($n$) 
\eqref{eq:neutron-weights} and BMA($p$) 
\eqref{eq:proton-weights}.
To assess the whole landscape, we apply   a local model averaging variant called BMA($n+p$), 
with local weights
\begin{equation}\label{eq:combined-weights}
\begin{split}
w_k(Z, N) & = w_k(n) H(N \geq N_{\beta}(Z))\\
&  + w_k(p) H(N < N_{\beta}(Z)),
\end{split}
\end{equation}
where $H(x)$ is  the Heaviside step function and $N_{\beta}(Z)$ is 
the neutron number corresponding 
to the average line of $\beta$-stability  defined as in Ref.~\cite{Kodama1971}.

\section{Results}\label{results}

The analysis of individual nuclear models' residuals in the context of theory developments  has been discussed in, e.g.,  Ref.~\cite{Neufcourt2018}. In this manuscript, we rather focus on quantified predictions of separation energies and drip lines, aiming to highlight the regions of the nuclear chart where the next-generation rare-isotope facilities may have the largest impact on theoretical modeling. The second goal is to  provide predictions for drip lines  with reliable uncertainties.

\subsection{Model mixing}

The   model weights obtained in both   BMA variants are listed 
in Table \ref{table:post-weights}.
We can see that BMA($n$) is well balanced between the models,
while the BMA($p$) is more selective. As discussed in Ref.~\cite{Neufcourt2020},
BMA($p$) heavily penalizes large deviations at single locations. 

By design, BMA($n+p$) retains the best of other two variants; it constitutes an innovative attempt for a
principled local model averaging, 
which we called for in previous works
\cite{Kej19,Neufcourt2020}.
Indeed, a model is arguably 
designed  to reproduce a particular phenomenon of interest, 
and while the idea of universality is appealing, 
a dogmatic extension of a model to a wider domain 
can dangerously amount to overfitting.
Local model averaging is particularly suited for situations where
desired accuracy is  high, and high-resolution effects must be taken into account to explain observations, in contrast to more qualitative descriptions.

In practice, all three BMA variants achieve a similar rms deviation ($S_{1n}\approx 302$\,keV, $S_{2n}\approx 453$\,keV, $S_{1p}\approx 410$\,keV, $S_{2p}\approx 438$\,keV),
using AME16-03 as an independent (extrapolative)
testing dataset
(see \cite{Neufcourt2020} for methodology details).

While the theoretical statistical foundations of a general local averaging framework 
are yet to be set, in our simplified setup it corresponds to the hypothesis
that neutron and proton separation energies are given by independent 
statistical models, which also matches our GP modeling assumption. 

The BMA weights can be used to assess the relative predictive power of the individual models corrected with the GP: UNEDF0, FRDM-2012 and SV-min reach highest evidence on the neutron-rich side, and FRDM-2012, HFB-24 and D1M perform the best on the proton-rich side.
Nevertheless, the relatively broad distribution of the weights suggests that no single model dominates.

\subsection{Landscape of nuclear existence}

Following Refs.~\cite{Neufcourt2019a,Neufcourt2020} we compute the probability  $p_{\rm ex}$  that  a given isotope is particle-bound, i.e., that  $S_{2p}>0$ for even-$Z$ nuclei, $S_{2n}>0$ for even-$N$ nuclei,   $S_{1p}>0$ for odd-$Z$ nuclei, and $S_{1n}>0$ for odd-$N$ nuclei. Formally, this quantity can be defined as:
\begin{equation}\label{pex}
p_{\rm ex}:=p(S_{1p/2p/1n/2n}^* > 0| S_{1p/2p/1n/2n}).
\end{equation}
Since the proton and neutron drip lines are well separated, one can write:
\begin{equation}\label{pex}
p_{\rm ex}=
p(S_{1p/2p}^* > 0| S_{1p/2p})\cdot p(S_{1n/2n}^* > 0| S_{1n/2n}),
\end{equation}
where $p(S_{1p/2p}^* > 0| S_{1p/2p})$ was obtained with BMA($p$) and $p(S_{1n/2n}^* > 0| S_{1n/2n})$ -- with BMA($n$).

The  drip line corresponds to $p_{\rm ex}=0.5$. Figure~\ref{fig:posterior-landscape} shows the posterior probability of existence $p_{\rm ex}$ for all nuclei in the nuclear landscape. The ranges of nuclear mass measurements and known nuclei are marked.
To provide a representative  example of discovery potential of  next-generation radioactive ion beam facilities, the figure  also shows the isotopes that will be accessible at the future  Facility for Rare Isotope Beams (FRIB) \cite{Glasmacher17,Sherrill}.

The FRIB production rates have been  estimated  with the \lisepp code \cite{LISE}. Production cross-sections for projectile fragmentation and fission reactions were obtained by using
the EPAX2.15 cross-section systematics \cite{Summerer} and the \lisepp 3EER Abrasion-Fission model \cite{LISEpp,LISEpp2}. FRIB rates and details of their calculations are available online \cite{fribrates}. In our estimates, we assumed  the  experimental limit for the confirmation of existence of an isotope to be 1 event/2.5 days.

For neutron-rich nuclei, FRIB will approach the  neutron drip line in the regions of neutron magic numbers. The magic nuclei are important for the r-process as they serve as  major bottlenecks in the synthesis of heavier elements.
In the region of  proton-rich nuclei, due
to the presence of the Coulomb barrier, relatively
long-lived, proton-unstable nuclei can exist beyond the proton
drip line~\cite{Pfutzner12,Neufcourt2020}. As seen
in Fig.~\ref{fig:posterior-landscape}, FRIB will reach  the uncharted  territory of many heavy proton-unstable nuclei. 

To accompany Fig.~\ref{fig:posterior-landscape}, 
we tabulate in \cite{SM} the calculated posterior predictions 
for particle separation energies 
for all drip-line nuclei with $0.1 < p_{ex} < 0.9$. 

\subsection{Neutron-rich nuclei}

\begin{figure*}[htb!]
\includegraphics[width=0.9\linewidth]{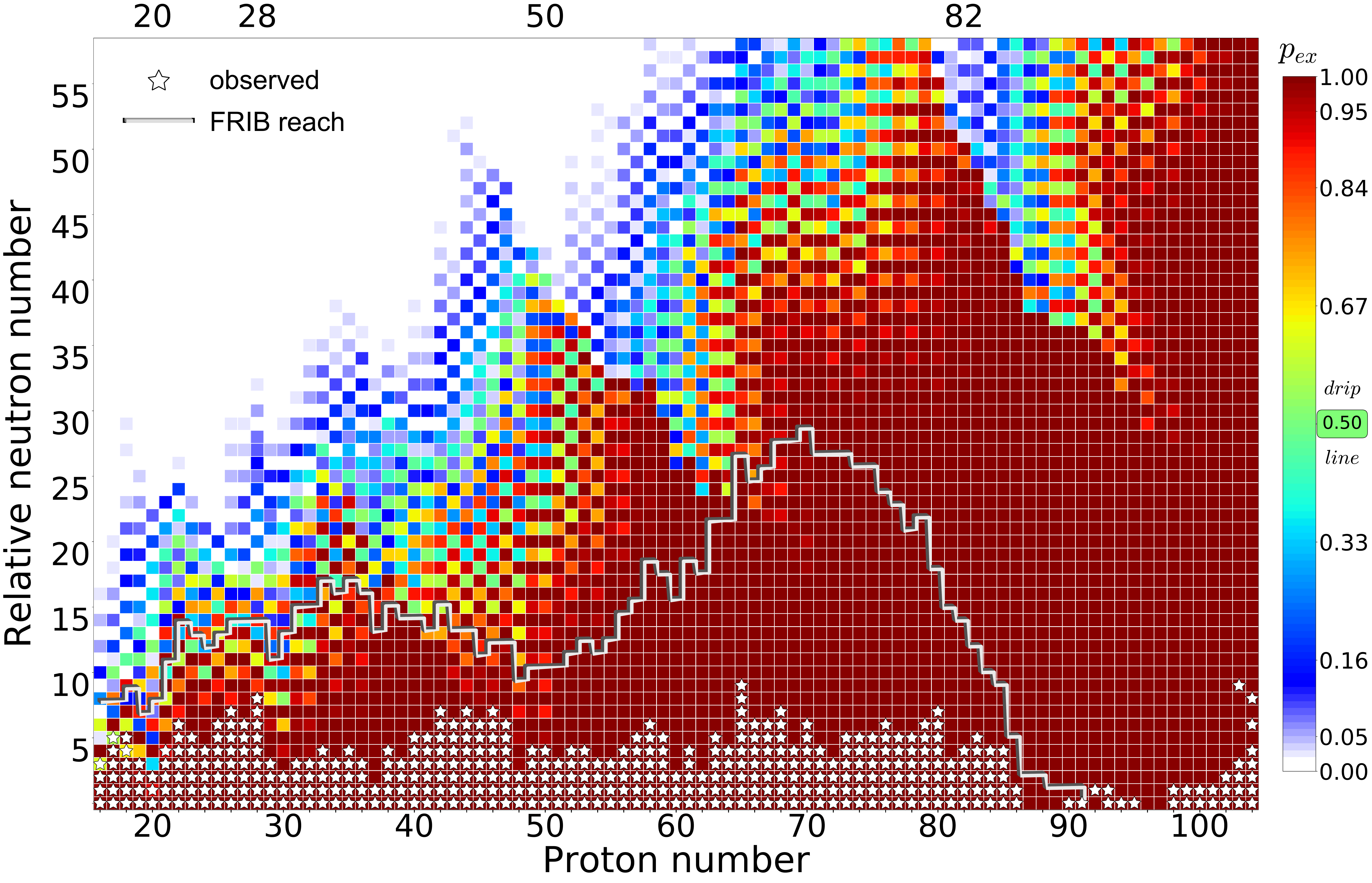}
\caption{
The quantified  separation energy landscape in the neutron drip-line region obtained with the BMA($n$) model averaging.
The color marks the ``probability of existence" $p_{\rm ex}$ of neutron-rich nuclei, i.e., the probability that  these nuclei are bound with respect to  neutron decay. 
For each proton number, $p_{\rm ex}$ is shown  along the isotopic chain versus
 the relative neutron number $N - N_0(Z)$, where
 $N_0(Z)$, listed in Table~\ref{tab:ref-even}, 
 is the neutron number of the heaviest isotope
 for which an experimental one- or two-neutron separation energy value is available.
The domain of nuclei that have been experimentally observed is marked by stars. 
To provide a realistic estimate of the discovery potential with  modern radioactive ion-beam facilities,
the isotopes within  FRIB's experimental reach are delimited by the shadowed solid line. See text for details. This figure (without the  FRIB range), in PDF format, can be downloaded from \cite{SM}.
}
\label{fig:neutron-dripline}
\end{figure*}

The quantified  separation energy landscape for  neutron rich nuclei, predicted in BMA($n$),  is displayed in   Fig.~\ref{fig:neutron-dripline}. To facilitate the presentation, the information for each isotope is given  relative to the neutron number 
$N_0$ of the heaviest neutron-bound isotope
 for which an experimental one- or two-neutron separation energy value is available.
The reference values of  $N_0(Z)$ are listed in Table~\ref{tab:ref-even}.
(For a similar diagram for proton-rich nuclei, see Ref.~\cite{Neufcourt2020}.)
To illustrate how to read Fig.~\ref{fig:neutron-dripline}, we consider the Ni isotopic chain.
The heaviest Ni isotope, for which mass has been measured is $^{73}$Ni \cite{Rahaman2007}; hence, $N_0(28)=45$. The stars at $Z=28$ indicate the isotopes $^{74-82}$Ni, which have been detected experimentally \cite{Sumikama2017}.
The nucleus $^{87}$Ni is expected to have $p_{\rm ex} < 50\%$, i.e.,  it is predicted to lie outside the one-neutron drip line. Because of pairing correlations, the two-neutron drip line for the Ni chain is shifted all the way to $N\approx 66$: the extremely neutron-rich isotope $^{92}$Ni is predicted to be the last bound isotope.

Figure~\ref{fig:neutron-dripline} also marks the reach of the FRIB facility, again as an example of what perhaps will be achievable experimentally.  According to our analysis, FRIB will reach the one-neutron drip line up to $Z=42$ (Mo) and will approach it again in the Sm-Gd region.
For the Ni chain, the current phase of FRIB is expected to produce meaningful data on $^{86-87}$Ni.
The use of  fragmentation reactions will  allow  to study the existence of nuclides in the region of $Z= 16-24$, where the crucial check for theoretical models is provided by studying the neutron stability of  $^{61}$Ca \cite{Neufcourt2019a}. 

As seen in Fig.~\ref{fig:neutron-dripline}, of particular importance for constraining theory are the existence data for $Z=28-30$,  $Z=42-48$, and $Z=64-66$. In all these cases, the one-neutron drip line is within experimental reach and theoretical uncertainties on the drip-line position are appreciable. The extension of mass measurements to more neutron-rich nuclei
in the Ca-Ni  and Cd-Sn regions will be of great value. Those can be carried out via the variety of methods, especially the time-of-flight technique \cite{Famiano2019} that can be applied to short-lived nuclides with  1-100 ms lifetimes.

\begin{table*}[htb!]
      \begin{ruledtabular}
     \caption{Table of the reference neutron numbers for even-Z nuclei
(used in Fig.~\ref{fig:neutron-dripline}).}
\label{tab:ref-even} 
\begin{tabular}{lccccccccccccccccccccccc}
\multicolumn{24}{c}{Even-$Z$}\\
$Z$:      & 16    & 18    & 20    & 22    & 24    & 26    & 28    & 30    & 32    & 34    & 36    & 38    & 40    & 42    & 44    & 46    & 48    & 50    & 52    & 54    & 56    & 58    & 60 \\
$N_0$:   & 29    & 30    & 37    & 35    & 40    & 42    & 45    & 52    & 54    & 57    & 61    & 64    & 67    & 69    & 73    & 77    & 83    & 85    & 88    & 92    & 93    & 93    & 100 \\
 \\[-4pt]
$Z$:       & 62    & 64    & 66    & 68    & 70    & 72    & 74    & 76    & 78    & 80    & 82    & 84    & 86    & 88    & 90    & 92    & 94    & 96    & 98    & 100   & 102   & 104   &  \\
$N_0$:   & 102   & 102   & 103   & 104   & 108   & 114   & 117   & 120   & 124   & 128   & 133   & 138   & 143   & 146   & 147   & 148   & 152   & 155   & 156   & 157   & 155   & 154   & \\\hline \\[-4pt]
\multicolumn{24}{c}{Odd-$Z$}\\
$Z$:     & 17    & 19    & 21    & 23    & 25    & 27    & 29    & 31    & 33    & 35    & 37    & 39    & 41    & 43    & 45    & 47    & 49    & 51    & 53    & 55    & 57    & 59 \\
$N_0$:  & 29    & 34    & 36    & 38    & 41    & 44    & 50    & 52    & 54    & 58    & 65    & 66    & 69    & 72    & 76    & 78    & 83    & 87    & 89    & 93    & 94    & 96 \\
 \\[-4pt]
$Z$:     & 61    & 63    & 65    & 67    & 69    & 71    & 73    & 75    & 77    & 79    & 81    & 83    & 85    & 87    & 89    & 91    & 93    & 95    & 97    & 99    & 101   & 103 \\
$N_0$: & 99    & 100   & 99    & 104   & 107   & 112   & 115   & 118   & 122   & 124   & 132   & 135   & 139   & 146   & 147   & 147   & 149   & 151   & 154   & 156   & 157   & 153
\end{tabular}%
      \end{ruledtabular}
      \end{table*}

\subsection{Number of particle-bound nuclei}

\begin{figure*}[htb!]
\includegraphics[width=0.8\linewidth]{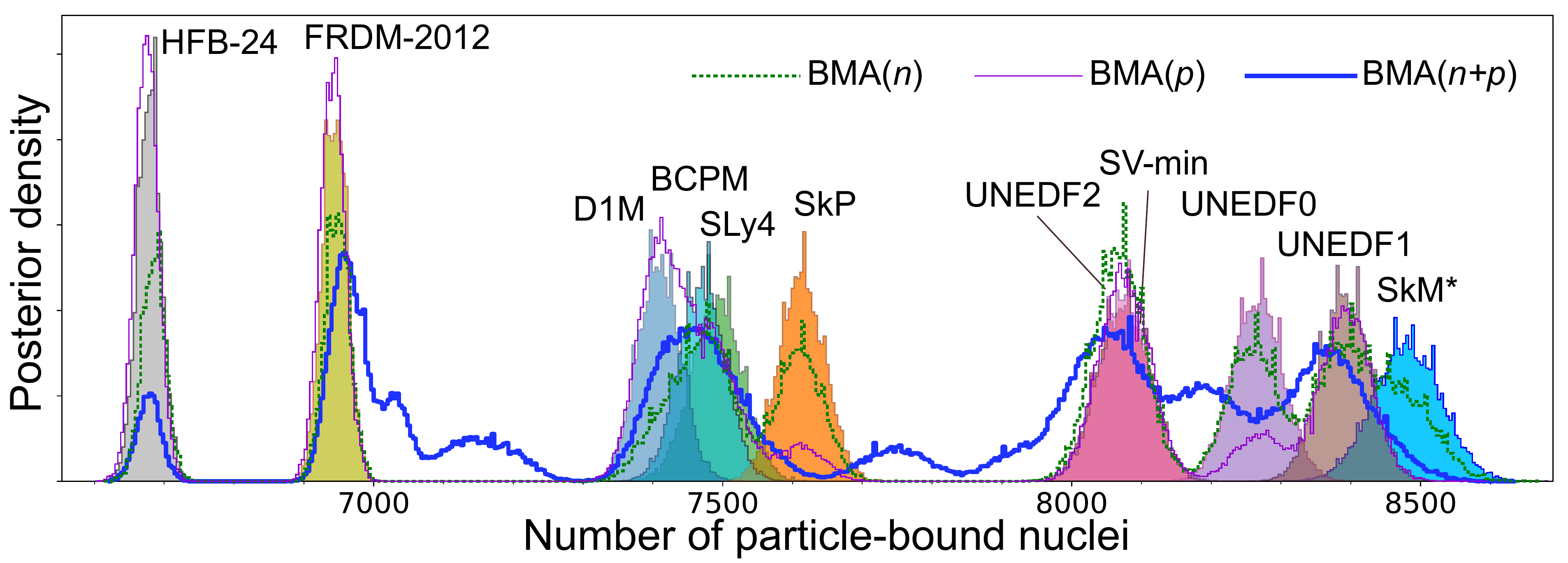}
\caption{
Posterior distributions of the number of particle-bound nuclei.
The histograms show the posterior densities
for each model: the peaks correspond successively to 
HFB-24, FRDM-2012, D1M, BCPM, SLy4, SkP, SV-min and UNEDF2,
UNEDF0, UNEDF1, and SkM$^*$.
The  lines shows the BMA posterior densities (multiplied by a constant factor  6.3 to facilitate presentation).
}
\label{fig:counts}
\end{figure*}

To estimate how many particle-bound nuclei  exist in the nuclear landscape,
we calculate the posterior distribution of the number of isotopes
with positive one- and two-nucleon separation energies.
We first produce such samples for each individual model,
which are then resampled into BMA posterior distributions.
These posterior distributions
are shown in Fig.~\ref{fig:counts}.

The number of nuclei with $Z,N\ge 8$ and $Z\leq 119$ predicted to be particle-bound by the individual models
range from 6600 (HFB-24) to 8600 (SkM$^*$).  This difference comes from the neutron-rich heavy nuclei for which the extrapolation uncertainty is very significant.
The BMA$(n)$ distribution has its average at 7765 ($\pm 590$ standard deviation), 
with median at 8032 and centered $95\%$ credibility interval $[6669, 8516]$.
The BMA$(p)$ distribution has its average at 7504 
($\pm 602$ standard deviation), 
with median at 7445 and centered $95\%$ credibility interval $[6661, 8425]$.

BMA($n+p$) amounts here to summing the number
of neutron-rich nuclei obtained from the BMA($n$) posterior distribution
and the number
of proton-rich nuclei obtained from the BMA($p$) posterior distribution -- hence the BMA($n+p$) distribution
is a convolution of BMA($n$) and BMA($p$),
which explains the smoothing effect seen in Fig.~\ref{fig:counts}.

Accordingly, the values obtained from BMA($n+p$) lie in between
with an average at 7708 ($\pm 534$ standard deviation)
median at 7785 and centered $95\%$ credibility interval $[6688, 8440]$.
It is  noticed that these bounds are tighter 
than those obtained with either BMA($n$) or BMA($p$).

Thus we can state without taking much risk that there should be between 6500 and 8500 stable nuclei based on the available mass data and models considered. While this result is consistent with the outcome of the earlier work \cite{Erl12a} employing uniform model mixing, the present study provides for the first time the detailed posterior distribution of the number of nuclei bound for each model. This represents a significant refinement of previous work that has been  allowed by our Bayesian statistical approach.

Figure~\ref{fig:counts} suggests that models can be clustered into 
three groups, where the more phenomenological ones 
yield the lowest number of particle-bound nuclei. Also,
it is worth noting that the models with similar and high weights (such as UNEDF0 and FRDM-2012) predict rather different numbers of particle bound-nuclei. This is not too surprising:  models tend to agree better in the domain of experimental data than at the location of the neutron drip line for the heaviest nuclei, where the available data allow only limited discrimination. It is expected that the future mass data on neutron-rich nuclei will provide more model selectivity.

\section{Conclusions}\label{conclusions}

By considering several global models and the most recent data on nuclear existence and masses, we applied novel Bayesian model averaging techniques to quantify the limits of the nuclear landscape.
We hope the drip-line estimates
as well as the specific predictions of one- and two-nucleon separation energies presented in this work will  guide experimental research at next-generation  rare isotope facilities. For instance, the  posterior predictions of particle separation energies of drip-line nuclei tabulated in \cite{SM}  can  be useful  when planning experiments aiming at establishing the existence of exotic isotopes. The related theoretical errors can guide  the uncertainty analysis for the r-process abundance studies.

As we emphasized in  previous studies~\cite{Neufcourt2019a,Neufcourt2020}, one should not expect that machine learning alone, however advanced, will somehow compensate for unknown systematic model deficiencies when extrapolating far away from the experimentally-established domain.  Indeed, since the range of our extrapolations is 2-3 times larger than the fitted range of the correlation effects, we can expect the GP correction to the predictions, apart for the shift $\mu$,  to be relatively limited. Consequently, in the unknown regions,  far extrapolations must rely on quality nuclear modeling. Therein, the honest evaluation of posterior predictive distributions is the key, i.e.,   the correction to the mean value  is of less importance compared to credibility intervals. In this respect, the GP extension to  nonzero $\mu$ as done in this work is perhaps more valuable than speculating about  a more elaborate GP tail model, which - if not substantiated by physics - would not offer any obvious advantages.

In our BMA calculations, we applied three model-mixing techniques. Two of them, the local models BMA($n$) and BMA($p$), have been informed by the specific data on extreme nuclei pertaining to very different domains. Namely, for BMA($n$) it is the existence of   neutron-rich isotopes with unknown  masses;  for BMA($p$) these are   $2p$ separation energies of five true $2p$ emitters.
The third global method BMA(n+p) retains locally each of these two variants on the part of the nuclear chart where it is, by design, expected to perform best.

According to our BMA($n+p$) analysis, the number of particle-bound nuclei with 
$Z, N \ge 8$ and $Z \le 119$  is  $7708\pm 534$. The results of the
individual models  shown in Fig.~\ref{fig:counts} show considerable spread, primarily due to the extrapolation uncertainty in the heavy neutron-rich region. This result underlines the fact that one should be very careful when trusting extrapolative predictions of any given model.

The extrapolations obtained in this study are timestamped. With the influx of new experimental data on existence and masses of exotic nuclei, and with new global mass models of high fidelity, the  quantified landscape of nuclear existence will gradually evolve.

\begin{acknowledgments}
Useful comments from Alexandra Gade are gratefully acknowledged.
Computational resources for statistical simulations 
were provided to L.N.
by the Institute for Cyber-Enabled Research at Michigan State University 
as well as Research Credits awarded by Google Cloud Platform. 
This material is based upon work supported by the U.S.\ Department of Energy, Office of Science, Office of Nuclear Physics under  award numbers DE-SC0013365 (Michigan State University), DE-SC0018083 (NUCLEI SciDAC-4 collaboration), and  DOE-NA0003885
(NNSA, the Stewardship Science Academic
Alliances program).
\end{acknowledgments}


%

\end{document}